\documentstyle[12pt]{article}
\newlength{\extraspace}
\newlength{\extraspaces}
\newcommand{\be}{\begin{equation}
\addtolength{\abovedisplayskip}{\extraspaces}
\addtolength{\belowdisplayskip}{\extraspaces}
\addtolength{\abovedisplayshortskip}{\extraspace}
\addtolength{\belowdisplayshortskip}{\extraspace}}
\newcommand{\ee}{\end{equation}}
\newcommand{\ba}{\begin{eqnarray}
\addtolength{\abovedisplayskip}{\extraspaces}
\addtolength{\belowdisplayskip}{\extraspaces}
\addtolength{\abovedisplayshortskip}{\extraspace}
\addtolength{\belowdisplayshortskip}{\extraspace}}
\newcommand{\ea}{\end{eqnarray}}
\newcommand{\nonu}{\nonumber \\[.5mm]}
\newcommand{\A}{&\!\!\!}
\newcommand{\slpartial}{\!\!\not\!\partial}

\begin{document}
\thispagestyle{empty}
\begin{flushright}
SIT--LP--02/05 \\
STUPP--02--166 \\
{\tt hep-th/0205178} \\
May, 2002
\end{flushright}
\vspace{7mm}
\begin{center}
{\large \bf Linearizing $N = 2$ Nonlinear Supersymmetry
} \\[20mm]
{\sc Kazunari Shima}${}^{\rm a}$\footnote{
\tt e-mail: shima@sit.ac.jp}, \
{\sc Yoshiaki Tanii}${}^{\rm b}$\footnote{
\tt e-mail: tanii@post.saitama-u.ac.jp} \
and \
{\sc Motomu Tsuda}${}^{\rm a}$\footnote{
\tt e-mail: tsuda@sit.ac.jp}
\\[5mm]
${}^{\rm a}${\it Laboratory of Physics, 
Saitama Institute of Technology \\
Okabe-machi, Saitama 369-0293, Japan} \\[3mm]
${}^{\rm b}${\it Physics Department, Faculty of Science, 
Saitama University \\
Saitama 338-8570, Japan} \\[20mm]
\begin{abstract}
We investigate for the $N = 2$ supersymmetry (SUSY) a relation 
between a vector supermultiplet of the linear SUSY and the 
Volkov-Akulov model of the nonlinear SUSY. We express component 
fields of the vector supermultiplet in terms of Nambu-Goldstone 
fermion fields at the leading orders in a SUSY invariant way, 
and show the vector nature of the U(1) gauge field explicitly. 
A relation of the actions for the two models is also discussed 
briefly. 
\end{abstract}
\end{center}

\newpage


Spontaneous breaking of supersymmetry (SUSY) gives rise to 
Nambu-Goldstone (N-G) fermions \cite{SS}, as shown 
in the breaking scheme of Fayet-Iliopoulos \cite{FI} 
and O'Raifeartaigh \cite{O}. 
N-G fermions can be characterized by means of the nonlinear 
realization of global SUSY in the Volkov-Akulov (V-A) model \cite{VA}. 
Their coupling to supergravity \cite{FNFDZ} under a local SUSY 
invariant way was investigated for the V-A model 
in the framework of the super Higgs mechanism \cite{DZ,SW}, 
in which N-G fermions are converted to the longitudinal components 
of spin-3/2 fields. 
\par
On the other hand, the connection between the V-A model 
of the nonlinear SUSY and linear supermultiplets became clear 
from the early work by many authors \cite{IK,R,UZ}: 
Indeed, in Ref.\ \cite{IK} the general relationship between linear 
and nonlinear realizations of global SUSY was established. 
In Ref.\ \cite{R} it was shown explicitly that the V-A model 
is related to a scalar supermultiplet of the linear SUSY 
of Wess and Zumino \cite{WZ1} by constructing irreducible 
and SUSY invariant relations on a scalar supermultiplet 
of the linear SUSY. 
The explicit connection between the V-A model and a scalar 
supermultiplet in two-dimensional spacetime is also discussed 
in Ref.\ \cite{UZ}. 
As for a vector supermultiplet, its relationship to the V-A model 
was studied in Ref.\ \cite{IK} in the context of the coupling of the V-A 
action to the gauge multiplet action with the Fayet-Iliopoulos 
$D$ term of the linear SUSY. 
\par
Recently, one of the authors proposed the superon-graviton 
model (SGM) based upon the SO(10) super-Poincar\'e algebra 
from a composite viewpoint of matters \cite{KS1,KS2}, 
which may be a most economical supersymmetric unified model 
beyond the standard model. 
The fundamental action of the SGM is the Einstein-Hilbert type 
one obtained from the geometrical arguments of the local GL(4,R) 
invariance under new nonlinear SUSY transformations of the SGM 
spacetime, where there exist fermionic degrees of freedom 
(N-G fermions) at every four-dimensional curved spacetime 
point \cite{KS2}. The expansion of the SGM action in terms 
of graviton and superons (N-G fermions) with spin-1/2 has 
a very complicated and rich structure \cite{ST1}; 
indeed, it is a highly nonlinear one which consists of 
the Einstein-Hilbert action of the general relativity, 
the V-A action and their interactions. 
Also, the SGM action is invariant under at least 
$[{\rm global\ nonlinear\ SUSY}] \otimes [{\rm local\ GL(4,R)}] 
\otimes [{\rm local\ Lorentz}] \otimes [{\rm global\ SO(N)}]$ 
as a whole \cite{ST2}, which is isomorphic to 
the global SO(N) super-Poincar\'e symmetry. 
\par
In the SGM the (composite) eigenstates of the {\it linear} 
representation of SO(10) super-Poincar\'e algebra which is composed 
of superons are regarded as all observed elementary 
particles at low energy except graviton \cite{KS1,KS2}. 
For deriving the low energy physical contents of the SGM action, 
it is important to linearize such a highly nonlinear theory. 
As a preliminary to do this, it is useful to investigate the 
linearization of the V-A model in detail. 
%
%
In this respect, in addition to the work by many authors \cite{IK,R,UZ}, 
we have explicitly shown in Ref.\ \cite{STT} that 
the $N = 1$ V-A model is related to the total action 
of a U(1) gauge supermultiplet \cite{WZ2} 
of the linear SUSY with the Fayet-Iliopoulos $D$ term indicating 
a spontaneous SUSY breaking. 
In the work of Ref.\ \cite{STT} it became clear that the 
representations of component fields of a U(1) gauge 
supermultiplet in terms of the N-G fermion fields indicate 
the axial vector nature of the U(1) gauge field. 
In order to see its vector nature at least, we have to investigate 
the linearization of the V-A model with an extended SUSY. 
\par
In this paper we restrict our attention to the $N = 2$ SUSY 
and discuss a connection between the V-A model and an $N = 2$ 
vector supermultiplet \cite{Fa} of the linear SUSY in 
four-dimensional spacetime. In particular, we show that for the 
$N = 2$ theory a SUSY invariant relation between component fields of 
the vector supermultiplet and the N-G fermion fields can be 
constructed by means of the method used in Ref.\ \cite{R} starting 
from an ansatz given below (Eq.\ (\ref{ansatz})). We also briefly 
discuss a relation of the actions for the two models. 
\par
Let us denote the component fields of an $N = 2$ U(1) gauge 
supermultiplet \cite{Fa}, which belong to representations of 
a rigid SU(2), as follows; 
namely, $\phi$ for a physical complex scalar field, 
$\lambda_R^i$ $(i = 1, 2)$ for two right-handed Weyl spinor fields 
and $A_a$ for a U(1) gauge field in addition to $D^I$ $(I = 1,2,3)$ 
for three auxiliary real scalar fields 
required from the mismatch of the off-shell degrees of freedom 
between bosonic and fermionic physical fields.\footnote{
Minkowski spacetime indices are denoted by $a, b, ... = 0, 1, 2, 3$, 
and the flat metric is $\eta^{ab}= {\rm diag}(+1, -1, -1, -1)$. 
Gamma matrices satisfy $\{ \gamma^a, \gamma^b \} = 2\eta^{ab}$ 
and we define $\gamma^{ab} = {1 \over 2}[\gamma^a, \gamma^b]$.}
$\lambda_R^i$ and $D^I$ belong to representations {\bf 2} and {\bf 3} 
of SU(2) respectively while other fields are singlets. 
By the charge conjugation we define left-handed spinor fields as 
$\lambda_{Li} = C \bar\lambda_{Ri}^T$. 
We use the antisymmetric symbols $\epsilon^{ij}$ and 
$\epsilon_{ij}$ ($\epsilon^{12} = \epsilon_{21} = +1$) to raise 
and lower SU(2) indices as $\psi^i = \epsilon^{ij} \psi_j$, 
$\psi_i = \epsilon_{ij} \psi^j$. 
\par
The $N=2$ linear SUSY transformations of these component fields 
generated by constant spinor parameters $\zeta_L^i$ are 
\ba
\delta_Q \phi \A = \A - \sqrt{2} \bar\zeta_R \lambda_L, \nonu
\delta_Q \lambda_{Li} 
\A = \A - {1 \over 2} F_{ab} \gamma^{ab} \zeta_{Li} 
- \sqrt{2} i \slpartial \phi \zeta_{Ri} 
+ i ( \zeta_L \sigma^I )_i D^I, \nonu
\delta_Q A_a \A = \A - i \bar\zeta_L \gamma_a \lambda_L 
- i \bar\zeta_R \gamma_a \lambda_R, \nonu
\delta_Q D^I \A = \A \bar\zeta_L \sigma^I \slpartial \lambda_L 
+ \bar\zeta_R \sigma^I \slpartial \lambda_R, 
\label{lsusy}
\ea
where $\zeta_{Ri} = C \bar\zeta_{Li}^T$, 
$F_{ab} = \partial_a A_b - \partial_b A_a$, and $\sigma^I$ are 
the Pauli matrices. The contractions of SU(2) indices 
are defined as 
$\bar\zeta_R \lambda_L = \bar\zeta_{Ri} \lambda_L^i$, 
$\bar\zeta_R \sigma^I \lambda_L 
= \bar\zeta_{Ri} (\sigma^I)^i{}_j \lambda_L^j$, etc. 
These supertransformations satisfy a closed off-shell 
commutator algebra 
\be
[ \delta_Q(\zeta_1), \delta_Q(\zeta_2)] 
= \delta_P(v) + \delta_g(\theta), 
\label{commutator}
\ee
where $\delta_P(v)$ and $\delta_g(\theta)$ are a translation and 
a U(1) gauge transformation with parameters 
\ba
v^a \A = \A 2i ( \bar\zeta_{1L} \gamma^a \zeta_{2L} 
- \bar\zeta_{1R} \gamma^a \zeta_{2R} ), \nonu
\theta \A = \A - v^a A_a + 2 \sqrt{2} \bar\zeta_{1L} \zeta_{2R} \phi 
- 2 \sqrt{2} \bar\zeta_{1R} \zeta_{2L} \phi^*. 
\label{u1}
\ea
Only the gauge field $A_a$ transforms under the U(1) gauge 
transformation 
\be
\delta_g(\theta) A_a = \partial_a \theta. 
\ee
\par
Although in our discussion on the derivation of 
the relation between the linear and nonlinear SUSY transformations 
we have not used a form of the action, it is instructive to consider 
a free action which is invariant under Eq.\ (\ref{lsusy}) 
\be
S_{\rm lin} = \int d^4 x \left[ \partial_a \phi \partial^a \phi^* 
- {1 \over 4} F^2_{ab} 
+ i \bar\lambda_R \!\!\not\!\partial \lambda_R 
+ {1 \over 2} (D^I)^2 - {1 \over \kappa} \xi^I D^I \right], 
\label{lact}
\ee
where $\kappa$ is a constant whose dimension is $({\rm mass})^{-2}$ 
and $\xi^I$ are three arbitrary real parameters satisfying 
$(\xi^I)^2 = 1$. The last term proportional to $\kappa^{-1}$ is an 
analog of the Fayet-Iliopoulos $D$ term in the $N=1$ theories \cite{FI}. 
The field equations for the auxiliary fields 
give $D^I = \xi^I / \kappa$ indicating a spontaneous SUSY breaking. 
\par
On the other hand, in the $N = 2$ V-A model \cite{BV} we have 
a nonlinear SUSY transformation law of the N-G fermion 
fields $\psi_L^i$ 
\be
\delta_Q \psi_L^i = {1 \over \kappa} \zeta_L^i 
- i \kappa \left( \bar\zeta_L \gamma^a \psi_L 
- \bar\zeta_R \gamma^a \psi_R \right) 
\partial_a \psi_L^i, 
\label{nlsusy}
\ee
where $\psi_{Ri} = C \bar\psi_{Li}^T$. This transformation 
satisfies off-shell the commutator algebra (\ref{commutator}) 
without the U(1) gauge transformation on the right-hand side. 
The V-A action invariant under Eq.\ (\ref{nlsusy}) reads 
\be
S_{\rm VA} = - {1 \over {2 \kappa^2}} \int d^4 x \; \det w, 
\label{vaact}
\ee
where the $4 \times 4$ matrix $w$ is defined by 
\be
w{^a}_b = \delta^a_b + \kappa^2 t{^a}_b, \qquad 
t{^a}_b = - i \bar\psi_L \gamma^a \partial_b \psi_L 
+ i \bar\psi_R \gamma^a \partial_b \psi_R. 
\ee
The V-A action (\ref{vaact}) is expanded in $\kappa$ as 
\ba
S_{\rm VA} \A = \A 
- {1 \over {2 \kappa^2}} \int d^4 x 
\left[ 1 + \kappa^2 t{^a}_a 
+ {1 \over 2} \kappa^4 (t{^a}_a t{^b}_b 
- t{^a}_b t{^b}_a) \right. \nonu
\A\A
\left. - {1 \over 6} \kappa^6 \epsilon_{abcd} \epsilon^{efgd} 
t{^a}_e t{^b}_f t{^c}_g 
- {1 \over 4!} \kappa^8 \epsilon_{abcd} \epsilon^{efgh} 
t{^a}_e t{^b}_f t{^c}_g t{^d}_h 
\right]. 
\label{vaactex}
\ea
\par
We would like to obtain a SUSY invariant relation between the 
component fields of the $N = 2$ vector supermultiplet and 
the N-G fermion fields $\psi^i$ at the leading orders of $\kappa$. 
It is useful to imagine a situation in which the linear SUSY 
is broken with the auxiliary fields having expectation values 
$D^I = \xi^I / \kappa$ as in the free theory (\ref{lact}). 
Then, we expect from the experience in the $N = 1$ cases 
\cite{IK,R,UZ} and the transformation law of the spinor 
fields in Eq.\ (\ref{lsusy}) that the relation should have a form 
\ba
\lambda_{Li} \A = \A i \xi^I (\psi_L \sigma^I)_i 
+ {\cal O}(\kappa^2), \nonu
D^I \A = \A {1 \over \kappa} \xi^I + {\cal O}(\kappa), \nonu
({\rm other\ fields}) \A = \A {\cal O}(\kappa). 
\label{ansatz}
\ea
Higher order terms are obtained such that the linear 
SUSY transformations (\ref{lsusy}) are reproduced by the nonlinear 
SUSY transformation of the N-G fermion fields (\ref{nlsusy}). 
\par
After some calculations we obtain the relation between the fields 
in the linear theory and the N-G fermion fields as 
\ba
\phi(\psi) \A = \A {1 \over \sqrt{2}} \, i \kappa \xi^I 
\bar\psi_R \sigma^I \psi_L 
- \sqrt{2} \kappa^3 \xi^I \bar\psi_L \gamma^a \psi_L 
\bar\psi_R \sigma^I \partial_a \psi_L \nonu
\A\A - {\sqrt{2} \over 3} \kappa^3 \xi^I \bar\psi_R \sigma^J \psi_L 
\bar\psi_R \sigma^J \sigma^I \!\!\not\!\partial \psi_R 
+ {\cal O}(\kappa^5), \nonu
\lambda_{Li}(\psi) \A = \A i \xi^I (\psi_L \sigma^I)_i 
+ \kappa^2 \xi^I \gamma^a \psi_{Ri} \bar\psi_R \sigma^I 
\partial_a \psi_L 
+ {1 \over 2} \kappa^2 \xi^I \gamma^{ab} \psi_{Li} 
\partial_a \left( \bar\psi_L \sigma^I \gamma_b \psi_L \right) \nonu
\A\A + {1 \over 2} \kappa^2 \xi^I ( \psi_L \sigma^J )_i 
\left( \bar\psi_L \sigma^J \sigma^I \!\!\not\!\partial \psi_L 
- \bar\psi_R \sigma^J \sigma^I \!\!\not\!\partial \psi_R \right) 
+ {\cal O}(\kappa^4), \nonu
A_a(\psi) \A = \A - {1 \over 2} \kappa \xi^I \left( 
\bar\psi_L \sigma^I \gamma_a \psi_L 
- \bar\psi_R \sigma^I \gamma_a \psi_R \right) \nonu
\A\A + {1 \over 4} i \kappa^3 \xi^I \biggl[ 
\bar\psi_L \sigma^J \psi_R \bar\psi_R \left( 2 \delta^{IJ} \delta_a^b 
- \sigma^J \sigma^I \gamma_a \gamma^b \right) \partial_b \psi_L \nonu
\A\A - {1 \over 4} \bar\psi_L \gamma^{cd} 
\psi_R \bar\psi_R \sigma^I \left( 2 \gamma_a \gamma_{cd} \gamma^b 
- \gamma^b \gamma_{cd} \gamma_a \right) \partial_b \psi_L 
+ (L \leftrightarrow R) \biggr] + {\cal O}(\kappa^5), \nonu
D^I(\psi) \A = \A {1 \over \kappa} \xi^I 
- i \kappa \xi^J \left( 
\bar\psi_L \sigma^I \sigma^J \!\!\not\!\partial \psi_L 
- \bar\psi_R \sigma^I \sigma^J \!\!\not\!\partial \psi_R \right) \nonu
\A\A + \kappa^3 \xi^J \biggl[ \bar\psi_L \sigma^I \psi_R \partial_a 
\bar\psi_R \sigma^J \partial^a \psi_L 
- \bar\psi_L \sigma^K \gamma^c \psi_L \biggl\{ 
i \epsilon^{IJK} \partial_c \bar\psi_L \!\!\not\!\partial \psi_L 
\nonu
\A\A - {1 \over 2} \partial_a \bar\psi_L 
\sigma^J \sigma^K \sigma^I \gamma_c \partial^a \psi_L 
+ {1 \over 4} \partial_a \bar\psi_L \sigma^J \sigma^I \sigma^K 
\gamma^a \gamma_c \!\!\not\!\partial \psi_L \biggr\} \nonu
\A\A - {1 \over 4} \bar\psi_L \sigma^K \psi_R \left\{ 
\partial_a \bar\psi_R \sigma^J \sigma^I \sigma^K \gamma^b \gamma^a 
\partial_b \psi_L 
- \bar\psi_R \left( 2 \delta^{IK} + \sigma^I \sigma^K \right) 
\sigma^J \Box \psi_L \right\} \nonu
\A\A + {1 \over 16} \bar\psi_L \gamma^{cd} \psi_R \left\{ 
\partial_a \bar\psi_R \sigma^J \sigma^I \gamma^b \gamma_{cd} 
\gamma^a \partial_b \psi_L 
+ \bar\psi_R \sigma^I \sigma^J \gamma^b \gamma_{cd} \gamma^a 
\partial_a \partial_b \psi_L \right\} \nonu
\A\A + (L \leftrightarrow R) \biggr] + {\cal O}(\kappa^5). 
\label{relation}
\ea
The transformation of the N-G fermion fields (\ref{nlsusy}) 
reproduces the transformation of the linear theory (\ref{lsusy}) 
except that the transformation of the gauge field 
$A_a(\psi)$ contains an extra U(1) gauge transformation 
\be
\delta_Q A_a(\psi) = - i \bar\zeta_L \gamma_a \lambda_L(\psi) 
- i \bar\zeta_R \gamma_a \lambda_R(\psi) + \partial_a X, 
\ee
where 
\be
X = {1 \over 2} i \kappa^2 \xi^I \bar\zeta_L \left( 
2 \delta^{IJ} - \sigma^{IJ} \right) \psi_R 
\bar\psi_R \sigma^J \psi_L + (L \leftrightarrow R). 
\ee
The U(1) gauge transformation parameter $X$ satisfies 
\be
\delta_Q(\zeta_1) X(\zeta_2) 
- \delta_Q(\zeta_2) X(\zeta_1) = - \theta, 
\ee
where $\theta$ is defined in Eq.\ (\ref{u1}). 
Due to this extra term the commutator of two supertansformations 
on $A_a(\psi)$ does not contain the U(1) gauge transformation 
term in Eq.\ (\ref{commutator}). 
This should be the case since the commutator on $\psi$ does not 
contain the U(1) gauge transformation term. 
For gauge invariant quantities like $F_{ab}$ the transformations 
exactly coincide with those of the linear SUSY. 
In principle we can continue to obtain higher order terms in the 
relation (\ref{relation}) following this approach. 
However, it will be more useful to use the $N=2$ superfield 
formalism \cite{GSW} as was done in Refs.\ \cite{IK,UZ,STT} 
for the $N=1$ theories. 
\par
We note that the leading terms of $A_a$ in Eq.\ (\ref{relation}) 
can be written as 
\be
A_a = - \kappa \xi^1 \bar\chi \gamma_5 \gamma_a \varphi 
+ i \kappa \xi^2 \bar\chi \gamma_a \varphi 
- {1 \over 2} \kappa \xi^3 \left( \bar\chi \gamma_5 \gamma_a \chi 
- \bar\varphi \gamma_5 \gamma_a \varphi \right) 
+ {\cal O}(\kappa^3), 
\ee
where we have defined Majorana spinor fields 
\be
\chi = \psi_L^1 + \psi_{R1}, \qquad 
\varphi = \psi_L^2 + \psi_{R2}. 
\ee
When $\xi^1 = \xi^3 = 0$, this shows the vector nature of the U(1) 
gauge field as we expected. 
\par
The relation (\ref{relation}) reduces to that of the $N = 1$ 
SUSY by imposing, e.g.  $\psi_L^2 = 0$. 
When $\xi^1 = 1$, $\xi^2 = \xi^3 = 0$, we find 
$\lambda_{L2} = 0$, $A_a = 0$, $D^3 = 0$ and that the relation 
between $(\phi, \lambda_{L1}, D^1, D^2)$ and $\psi_L^1$ 
becomes that of the $N=1$ scalar supermultiplet obtained in 
Ref.\ \cite{R}. 
\footnote{
$N = 1$ scalar supermultiplet is also obtained 
by adopting $\xi^1 = 1 = \xi^2$, $\xi^3 = 0$.}
When $\xi^1 = \xi^2 = 0$, $\xi^3 = 1$, on the other hand, 
we find $\lambda_{L1} = 0$, $\phi = 0$, $D^1 = D^2 = 0$ 
and that the relation between $(\lambda_{L2}, A_a, D^3)$ and 
$\psi_L^1$ becomes that of the $N=1$ vector 
supermultiplet obtained in Refs.\ \cite{IK,STT}. 
\par
In our above discussion on the derivation of the result (\ref{relation}) 
we have not used a form of the action. 
However, as for the relation between the free linear SUSY action 
$S_{\rm lin}$ in Eq.\ (\ref{lact}) and the V-A action 
$S_{\rm VA}$ in Eq.\ (\ref{vaact}), 
we have explicitly shown that $S_{\rm lin}$ indeed coincides 
with the V-A action $S_{\rm VA}$ at least up to and including 
$O(\kappa^0)$ in Eq.\ (\ref{vaactex}) by substituting 
Eq.\ (\ref{relation}) into the linear action (\ref{lact}). 
We consider this is a strong indication of the all order 
coincidence between the actions (\ref{lact}) and (\ref{vaact}) 
from the experience in the $N = 1$ cases \cite{IK,R,UZ,STT}. 
\par
Finally we summarize our results. In this paper we have constructed 
the SUSY invariant relation between the component fields 
of the $N = 2$ vector supermultiplet and the N-G fermion fields 
$\psi_L^i$ at the leading orders of $\kappa$. 
We have explicitly showed that the U(1) gauge field $A_a$ 
has the vector nature in terms of the N-G fermion fields 
in contrast to the models with the $N = 1$ SUSY \cite{STT}. 
The relation (\ref{relation}) contains three arbitrary real 
parameters $\xi^I/\kappa$, which can be regarded as the vacuum 
expectation values of the auxiliary fields $D^I$. 
When we put $\psi_L^2 = 0$, the relation reduces to that of 
the $N = 1$ scalar supermultiplet or that of the $N = 1$ vector 
supermultiplet depending on the choice of the parameters $\xi^I$. 
We have also briefly discussed that the free action $S_{\rm lin}$ 
in Eq.\ (\ref{lact}) with the Fayet-Iliopoulos $D$ term 
reduces to the V-A action $S_{\rm VA}$ in Eq.\ (\ref{vaact}) 
at least up to and including $O(\kappa^0)$ by subsituting 
the relation (\ref{relation}) into the linear action (\ref{lact}). 
From the results in this Letter we anticipate 
the equivalence of the action of $N$-extended standard 
supermultiplets to the $N$-extended V-A action 
of a nonlinear SUSY, which is favorable for the SGM scenario.

\vspace{10mm}

\noindent {\Large{\bf Acknowledgements}}

\vspace{3mm}

The work of Y.T. is supported in part by the Grant-in-Aid from the 
Ministry of Education, Culture, Sports, Science and Technology, 
Japan, Priority Area (\#707) ``Supersymmetry and Unified Theory 
of Elementary Particles''. 
The work of M.T. is supported in part by the High-Tech research 
program of Saitama Institute of Technology. 


%
\newcommand{\NP}[1]{{\it Nucl.\ Phys.\ }{\bf #1}}
\newcommand{\PL}[1]{{\it Phys.\ Lett.\ }{\bf #1}}
\newcommand{\CMP}[1]{{\it Commun.\ Math.\ Phys.\ }{\bf #1}}
\newcommand{\MPL}[1]{{\it Mod.\ Phys.\ Lett.\ }{\bf #1}}
\newcommand{\IJMP}[1]{{\it Int.\ J. Mod.\ Phys.\ }{\bf #1}}
\newcommand{\PR}[1]{{\it Phys.\ Rev.\ }{\bf #1}}
\newcommand{\PRL}[1]{{\it Phys.\ Rev.\ Lett.\ }{\bf #1}}
\newcommand{\PTP}[1]{{\it Prog.\ Theor.\ Phys.\ }{\bf #1}}
\newcommand{\PTPS}[1]{{\it Prog.\ Theor.\ Phys.\ Suppl.\ }{\bf #1}}
\newcommand{\AP}[1]{{\it Ann.\ Phys.\ }{\bf #1}}

\end{document}